\documentclass{PoS}

\title{Multiple-channel generalization of Lellouch-L\"uscher formula}

\ShortTitle{Multiple-channel generalization of Lellouch-L\"uscher formula}

\author{\speaker{Maxwell T. Hansen}\footnote{This work was supported in part by the U.S. Department of Energy under Grant No. DE-FG02-96ER40956}\\
         Physics Department, University of Washington, Seattle, WA 98195-1560, USA\\
        Fermilab Theory Group, PO Box 500, MS 106, WH3E, Batavia, IL 60510, USA\footnote{Current address.}\\
        E-mail: \email{mth28@uw.edu}}

\author{Stephen R. Sharpe\\
        Physics Department, University of Washington, Seattle, WA 98195-1560, USA\\
        E-mail: \email{srsharpe@uw.edu}}

\usepackage[sort&compress,numbers]{natbib}

\newcommand{\MeV}{\mathop{\rm MeV}\nolimits}

\renewcommand{\Re}{\mathrm{Re\,}}

\abstract{We describe a generalization of the Lellouch-L\"uscher formula to the case of multiple strongly-coupled decay channels. As in the original formula, our final result is a relation between weak matrix elements in finite and infinite volumes. Our extension is limited to final states with two scalar particles, with center of mass energies below the lowest three- or four-particle threshold. Otherwise the extension is general, accommodating any number of channels, arbitrary strong coupling between channels, as well as any form of weak decay operators in the matrix elements. Among many possible applications, we emphasize that this is a necessary first step on the way to a lattice-QCD calculation of weak decay rates for $D\to\pi\pi$ and $D\to K \overline K$. Our results allow for arbitrary total momentum and hold for degenerate or non-degenerate particles.
}

\FullConference{The 30th International Symposium on Lattice Field Theory\\
                 June 24 - 29,  2012\\
                 Cairns, Australia}

\begin{document}

\section{Introduction}

Lattice calculations of weak decay amplitudes for processes with multiple hadrons in the final state are more difficult than calculations for leptonic or semi-leptonic decays. One of the many challenges faced is that finite volume effects are more important in the former case.

For the simplest hadronic decay, \(K \rightarrow \pi \pi\), the finite volume effects were worked out by L\"uscher and Lellouch over a decade ago \cite{Luescher:1986n1,Luescher:1986n2, Luescher:1991n1, Luescher:1991n2, Lellouch:2000}. This work, together with a great deal of additional development in both formalism and computational techniques, has allowed for significant progress towards a first principles determination of these weak matrix elements \cite{Blum:2011,Blum:2011ng}. Specifically, results for the isospin-two sector with a complete error budget are now available~\cite{Blum:2011ng}, and the more challenging isospin-zero sector is expected to follow within the next few years.

We are thus led to consider what information lattice calculations might eventually offer concerning heavier meson decays. For example, LHCb and CDF each recently reported CP-violation in the difference of CP-asymmetries for \(D^0 \rightarrow \pi^- \pi^+\) and \(D^0 \rightarrow K^- K^+\)~\cite{LHCb:2011, Collaboration:2012qw}. Lattice simulation is the only known method to check consistency of these results with the Standard Model. Therefore, though such a calculation is clearly very challenging, it is natural to begin investigating how it might be done. In particular, in contrast to the \(K \rightarrow \pi \pi\) case, the finite volume effects for decays of heavier mesons are not yet well understood. Our aim is to take a first step towards improving this understanding. Details can be found in the longer publication, Ref.~\cite{Hansen:2012tf}.

Specifically we show that, if one can ignore all but two-particle channels, then a generalization of the work of Lellouch and L\"uscher allows for a first principles calculation of \(D\) decay amplitudes from the finite volume spectrum obtained via lattice simulation. Incorporating multiple strongly coupled two-particle channels is a precursor to the more difficult challenge of also coupling in four and higher particle states. Indeed such states, for example four pions, are expected to be important at the \(D\) mass (\(M_{D^0}=1865 \MeV\)).

Our discussion begins with a statement of the multiple-channel quantization condition. This was obtained using a field theoretic derivation in Ref.~\cite{Hansen:2012tf} and is a necessary first step before deriving the Lellouch-L\"uscher formula. We note that there have been a large number of recent papers studying the generalization of the L\"uscher quantization condition to multiple two-body channels and assessing its utility~\cite{Liu:2005kr,Lage:2009,Bernard:2010,Doering:2011, Aoki:2011}. After giving the quantization condition, we then show how it may be used to derive the generalized Lellouch-L\"uscher formula, which allows one to extract amplitudes for decays into multiple, strongly coupled, two-particle states.

\section{Generalized L\"uscher quantization condition}

In this section we state, without derivation, our extension of L\"uscher's quantization condition \cite{Luescher:1986n1,Luescher:1986n2, Luescher:1991n1, Luescher:1991n2}, to the case of multiple, strongly-coupled channels. As emphasized above, the extension allows for any number of coupled channels, each containing two scalar particles. 

Throughout the article we take finite volume to mean a finite, cubic spatial volume of extent \(L\) with periodic boundary conditions. We take \(L\) large enough so that exponentially suppressed corrections can be ignored, and take the (Minkowski) time direction to be infinite.\footnote{Lattice simulations are performed with Euclidean time, and the spectrum is found from the exponential decay of correlators. In the corresponding Minkowski theory, the same spectrum is given by energy poles in momentum-space correlators. For our discussion the latter description turns out to be more convenient.} Although lattice simulation is the target of our investigation, we assume here that discretization errors are small and controlled and therefore use continuum field theory (zero lattice spacing).

The kinematic variables which enter our discussion are the total energy \(E\), the total momentum
\begin{equation}
\vec P = \frac{2 \pi \vec n_P}{L}\ \ \ \ \ \ (\vec n_P \in \mathbb{Z}^3) \,,
\end{equation}
and the center of mass (CM) frame energy \(E^* = \sqrt{E^2 - \vec P^2}\). The condition that only two-particle channels are open is effected by requiring that \(E^*\) sit below any higher particle state that is coupled to the sector of interest. In the case that a symmetry prevents even/odd coupling, this means taking \(E^*\) below the lowest four-particle threshold. 

The quantization condition, which we derive in Ref.~\cite{Hansen:2012tf} by following Ref.~\cite{Kim:2005}, is of the form
\begin{equation}
\label{eq:sec}
\Delta^{\mathcal M}(L,E^*, \vec P) = 0\,.
\end{equation}
Here \(\mathcal M\) represents the infinite volume scattering amplitudes and is defined explicitly in Eq.~(\ref{eq:mdef}) below. The explicit form of \(\Delta^{\mathcal M}\) is also given later in this section [Eq.~(\ref{eq:expsec})]. To understand the utility of Eq.~(\ref{eq:sec}) one should first suppose a theory of scalar particles, for which all two-to-two scattering amplitudes are known. Then, at fixed \(\{L, \vec P\}\), \(\Delta^{\mathcal M}\) becomes a known function of a single variable, \(E^*\). The content of Eq.~(\ref{eq:sec}) is that one should search this function for all roots, \(E_k^*\) (\(k=1,2,3,\cdots\)). These then give the energy spectrum of the finite volume theory. 

We comment that this description is actually the reverse of the most common case, in which one has energy levels from a lattice simulation and wants to determine physical scattering amplitudes. In this case one parametrizes the scattering amplitudes at fixed energy, \(E^*_0\), with a number \(K\) of real unknowns. Next one must determine \(K\) sets \(\{L_1, \vec P_1\}, \cdots \{L_K, \vec P_K\}\) which put \(E^*_0\) in the simulated spectrum. In this way one can produce \(K\) independent equations with the form of Eq.~(\ref{eq:sec}) and solve for the scattering amplitude. At the end of this section, we describe this method in detail for a particular toy model.

We now turn to a precise description of the scattering amplitudes \(\mathcal M\). We assume a total of \(N\) open two-particle channels, labeled \(i, j = 1, \cdots, N\). Then the two-to-two scattering amplitudes at fixed \(E^*\) depend only on (a) the particle content of the in and out-state and (b) the relative direction of motion between incoming and outgoing particles in the CM frame. It is convenient to give the scattering amplitude redundant dependence by including both incoming and outgoing CM frame directions of motion, labeled \(\hat k^*\) and \(\hat k^{*'}\) respectively. The amplitude may then be written 
\begin{equation}
\label{eq:mdef}
\mathcal M_{ij}(\hat k^*, \hat k^{*'}) = 4 \pi \mathcal M_{ij; \ell_1, m_1; \ell_2, m_2}  Y_{\ell_1, m_1}(\hat k^*) Y^*_{\ell_2, m_2}(\hat k^{*'})  \,,
\end{equation}
where a sum over \(\ell_1, m_1, \ell_2, m_2\) is implied. We comment that \(\mathcal M_{ij; \ell_1, m_1; \ell_2, m_2}\) is not diagonal in channel (\(ij\)) space but is diagonal in angular momentum (\(\ell_1, m_1; \ell_2, m_2\)) space:
\begin{equation}
\mathcal M_{ij; \ell_1, m_1; \ell_2, m_2} = \mathcal M_{ij}^{\ell_1,m_1} \delta_{\ell_1 \ell_2} \delta_{m_1 m_2} \,
\end{equation} 
(no sum). This is due to the rotational invariance of the infinite volume theory.

As a specific example, consider a toy model with three types of scalar particles: pions (with mass \(M_\pi\)) and kaons and anti-kaons (degenerate with mass \(M_K\)). Suppose, as is true for the isospin zero sector of physical pions and kaons, that the two pion (\(\pi \pi\)) and kaon anti-kaon (\(K \overline K\)) states only couple strongly to each other and also to states with four or more particles. It follows that, if we assume unphysical masses which satisfy \(2 M_K < 4 M_\pi\) and we also require that \(E^*\) lies below the four pion threshold, then the only open channels are \(\pi \pi\) and \(K \overline K\). In this case \(N=2\) with \(i=1\) for the pions and \(i=2\) for the kaons. Then, for example, \(\mathcal M_{12}\) is the \(K \overline K \rightarrow \pi \pi\) scattering amplitude.

Returning to arbitrary two-particle channels, the infinite dimensional matrix \(\mathcal M_{ij; \ell_1, m_1; \ell_2, m_2}\) is one of the two ingredients needed to define \(\Delta^{\mathcal M}\). The other is a kinematically determined matrix \(F\) which acts on the same space and is given by
\begin{equation}
F_{jk;\ell_1, m_1;\ell_2, m_2} \equiv \delta_{jk} \eta_j \left[\frac{\Re x_j}{4 L E^*} \delta_{\ell_1 \ell_2} \delta_{m_1 m_2} + \frac{i}{2 \pi E L} \sum_{\ell, m} x_j^{-\ell} \mathcal Z^P_{\ell,m}[1;x_j^2] \int d \Omega Y^*_{\ell_1,m_1} Y^*_{\ell,m} Y_{\ell_2,m_2} \right] \,,
\end{equation}
where \(x_j\) is the solution to
\begin{equation}
\frac{LE^*}{2 \pi} = \sqrt{x_j^2 + \left [ \frac{L M_{j,1}}{2\pi} \right ]^2} +  \sqrt{x_j^2 + \left [ \frac{L M_{j,2}}{2\pi} \right ]^2}
\end{equation}
with \(M_{j,1}\) and \(M_{j,2}\) equal to the two-particle masses of the \(j\)th channel. \(\mathcal Z^P_{\ell m}\) is a generalized zeta function defined in Ref.~\cite{Davoudi:2011md} and \(\eta = 1/2\) for identical and \(1\) for non-identical particles. We comment that, in contrast to \(\mathcal M\), \(F\) is diagonal in channel space but not diagonal in angular momentum space. This is due to the breaking of rotational symmetry by the finite volume condition.

We now state, without proof, the explicit form of our secular equation (\ref{eq:sec}):
\begin{equation}
\label{eq:expsec}
\Delta^{\mathcal M}(L, E^*, \vec P) \equiv \det(F^{-1} + i \mathcal M) = 0\,.
\end{equation}
This result agrees with that found in Ref.~\cite{Briceno:2012yi}. It is also agrees with the earlier work of Ref.~\cite{Bernard:2010} in the limiting case of only \(s\)-wave scattering and \(\vec P=0\).

We next note that Eq.~(\ref{eq:expsec}) is only practically useful in the case that the scattering amplitude is negligible above some \(\ell_{max}\) (\(\mathcal M_{ij}^{\ell>\ell_{max}, m}=0\)). In this case one can show, by extension of a proof in Ref.~\cite{Kim:2005}, that \(F\) may also be truncated with no additional approximation. This is nontrivial since \(F\) is not diagonal in angular momentum. However, the projection in \(\mathcal M\) turns out to be enough.

We now return to our toy model of pions and kaons and incorporate the additional assumption that the \(s\)-wave scattering amplitudes dominate (\(\ell_{max}=0\)). In this case Eq.~(\ref{eq:expsec}) takes on a particularly simple form:
\begin{equation}
\label{eq:sectwo}
\Delta^{\mathcal M}(L,E^*,\vec P) \equiv \det \! \Bigg [
\left( \begin{array}{ccc}
\! \! \big[F_1^s(L,E^*,\vec P)\big]^{-1}  & 0 \! \!\\
\! \! 0  & \big[F_2^s(L,E^*,\vec P)\big]^{-1} \! \! \end{array} \right) + i 
\left( \begin{array}{ccc}
\mathcal M^s_{1 \rightarrow 1}(E^*)  & \mathcal M^s_{2 \rightarrow 1}(E^*)\\
\mathcal M^s_{1 \rightarrow 2}(E^*)  & \mathcal M^s_{2 \rightarrow 2}(E^*) \end{array} \right) 
\Bigg ] = 0 \,,
\end{equation}
where all entries are now numbers and all functional dependence has been made explicit. 

We conclude this section by describing the extraction of scattering amplitudes in the \(s\)-wave \(\pi \pi\)-\(K \overline K\) model. Due to unitarity and symmetry of the \(S\)-matrix, the scattering amplitudes for this model are given at any fixed CM energy, \(E^*_0\), by three real parameters. For an explicit parametrization see Ref.~\cite{Hansen:2012tf}. One must therefore determine, from lattice simulation, three sets \(\{L_1, \vec P_1\}, \{L_2, \vec P_2\}, \{L_3, \vec P_3\}\) which put \(E^*_0\) in the spectrum. These may then be substituted into Eq.~(\ref{eq:sectwo}) to deduce three independent equations, which constrain the three unknowns and thus determine \(\mathcal M^s_{i \rightarrow j}(E^*_0)\).

\section{Generalized Lellouch-L\"uscher formula}

In this section we show how the quantization condition of the previous section may be used to derive the generalization of the Lellouch-L\"uscher formula, relating weak matrix elements in finite and infinite volume. We restrict ourselves, for the entire section, to the \(s\)-wave \(\pi \pi\)-\(K \overline K\) model introduced above, and therefore only reference the specific quantization condition (\ref{eq:sectwo}). The arguments given here can be easily generalized to any number of two-scalar-particle states.

We begin with a precise statement of the problem. We introduce an operator, \(\mathcal H_W(x)\), which weakly couples the \(\pi \pi\) and \(K \overline K\) to an otherwise non-interacting one particle state, which we suggestively label \(D\). Our aim is to derive a formula which takes as input \(\mathcal M\) and also the finite volume matrix elements
\begin{equation}
M_{D \rightarrow n} \equiv \langle n \vert \mathcal H_W(0) \vert D \rangle \,,
\end{equation}
and gives from this the infinite volume decay amplitudes
\begin{equation}
A_{D \rightarrow \pi\pi} \equiv \langle \pi \pi \vert \mathcal H_W(0) \vert D \rangle \ \ \ \ \ \ \ \ \ \ A_{D \rightarrow K \overline K} \equiv \langle K \overline K \vert \mathcal H_W(0) \vert D \rangle \,.
\end{equation}
Because our quantization condition is only valid below the four-pion threshold, we require \(M_D < 4 M_\pi \), where \(M_D\) is the \(D\) mass. This, together with \(2 M_K < 4 M_\pi\), means that our result is not an honest description of physical \(D\) decay. What we present here is a necessary first step towards that challenge.

The derivation proceeds by incorporating the weak Hamiltonian into the original theory via
\begin{equation}
\label{eq:modham}
\mathcal H(x) \longrightarrow \mathcal H(x) + \lambda \mathcal H_W(x) \,,
\end{equation}
where \(\lambda\) is a real parameter that we can freely vary and, in particular, make arbitrarily small. The new Hamiltonian changes the scattering amplitudes and therefore also changes the form of the quantization condition:
\begin{equation}
\Delta^{\mathcal M + \Delta \mathcal M}(L,E^*,\vec P)=0 \,.
\end{equation}
The leading order (in \(\lambda\)) form of the change to the amplitude \(\Delta \mathcal M\) is proportional to the matrix
\[ \left( \begin{array}{ccc}
A_{\pi\pi \rightarrow D} A_{D \rightarrow \pi \pi}\ \ \  &\ \ \  A_{K \overline K \rightarrow D} A_{D \rightarrow \pi \pi}  \\[5pt]
A_{\pi\pi \rightarrow D} A_{D \rightarrow K \overline K}\ \ \  &\ \ \  A_{K \overline K \rightarrow D} A_{D \rightarrow K \overline K} \end{array} \right) \,.\] 
This change is due to the diagram shown in Figure 1 in which the two scalar particles of the in-state combine to form a virtual \(D\) which then decays into the two scalar particles of the out-state.

\begin{figure}
\centering
\includegraphics{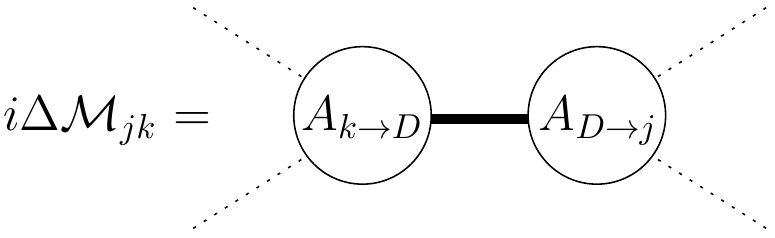}
\caption{The diagram giving rise to the amplitude perturbation \(\Delta \mathcal M\).}
\end{figure}

We now suppose that \(\{L_k, \vec P_k\}\) are chosen so that, in the absence of any weak interaction, \(M_D\) is in the two-particle spectrum:
\begin{equation}
\Delta^{\mathcal M}(L_k,E^*=M_D, \vec P_k)=0 \,.
\end{equation}
Here \(k=1,2,\cdots\) labels the different solutions to the quantization condition at \(E^*=M_D\). Now comes the key point: When the modified Hamiltonian [Eq.~(\ref{eq:modham})] is used, then the values \(L_k,M_D,\vec P_k\) no longer satisfy the quantization condition. However, by using leading order degenerate perturbation theory, one can correct the energies to restore the condition's validity. We conclude that
\begin{equation}
\label{eq:pertquant}
\Delta^{\mathcal M + \Delta \mathcal M}(L_n,E^*=M_D \pm \lambda L^3 \vert M_{D \rightarrow n} \vert, \vec P_n)=0 \,.
\end{equation}

Actually Eq.~(\ref{eq:pertquant}) is only valid through order \(\lambda\). Indeed, it is by expanding in \(\lambda\) and requiring that the linear coefficient vanish that we reach our generalization of the Lellouch-L\"uscher formula. The final result may be written as
\begin{equation}
\label{eq:LL}
\vert C_\pi(L_n, \vec P_n) A_{D \rightarrow \pi \pi} + C_K(L_n, \vec P_n) A_{D \rightarrow K \overline K} \vert = \vert M_{D \rightarrow n} \vert \,.
\end{equation}  
This is the main result of this section. The coefficients \(C_\pi\) and \(C_K\) depend on \(M_D\), \(\{L_n , \vec P_n\}\), \(\mathcal M(E^* = M_D)\) and \(d \mathcal M(E^*=M_D)/d E^*\). We direct the reader to Ref.~\cite{Hansen:2012tf} for their specific forms.\footnote{The notation of \cite{Hansen:2012tf} is slightly different, and is related to the notation in this article by
\[
C_\pi \equiv \vert \mathcal C_{M^2} \vert^{-1/2} \sqrt{q_1^* \eta_1} \left[c_1 e^{- i \delta_\alpha} c_\epsilon - c_2 e^{-i \delta_\beta} s_\epsilon \right] \,,
\]
\[
C_K \equiv \vert \mathcal C_{M^2} \vert^{-1/2} \sqrt{q_2^* \eta_2} \left[ c_1 e^{- i \delta_\alpha} s_\epsilon + c_2 e^{- i \delta_\beta} c_\epsilon \right] \,.
\]
All parameters appearing on the right hand side are defined in \cite{Hansen:2012tf}. Specifically \(\mathcal C_{M^2}\) is defined in Eq.~(88),  \(q^*\) in Eq.~(34), \(c_1\) and \(c_2\) in Eq.~(91) and \(\delta_\alpha\), \(\delta_\beta\), \(c_\epsilon\) and \(s_\epsilon\) in Eq.~(48).} 

To understand the use of Eq.~(\ref{eq:LL}) one need only recall that the quantization condition Eq.~(\ref{eq:sectwo}) may be used to determine the scattering amplitude in the region of \(E^*=M_D\). To get the amplitude one needs at least three sets \(\{L_1, \vec P_1\}\), \(\{L_2, \vec P_2\}\), \(\{L_3, \vec P_3\}\) which put \(M_D\) in the spectrum. Indeed one also needs three sets which put a slightly higher energy in the spectrum, in order to extract the derivative. Since three volume-momentum values are needed anyway, it is natural to calculate the corresponding three values of \(\vert M_{D \rightarrow n} \vert\) (by lattice simulation) and also the coefficients \(C_\pi\) and \(C_K\) (by substituting determined quantities into the equations of Ref.~\cite{Hansen:2012tf}). This results in three independent equations of the form
\begin{equation}
\label{eq:threeeqA}
\vert (\mathrm{known\ number}) A_{D \rightarrow \pi \pi} + (\mathrm{known\ number}) A_{D \rightarrow K \overline K} \vert = (\mathrm{known\ number}) \,.
\end{equation}

Indeed it turns out that three independent equations are precisely what is needed to determine the weak decay amplitudes. To see this one must first show, as we do in Ref.~\cite{Hansen:2012tf}, that \(A_{D \rightarrow \pi \pi}\) and \(A_{D \rightarrow K \overline K}\) may be written as a known linear combination of two real numbers, which we call \(v_1\) and \(v_2\).\footnote{For weak operators that violate time reversal (T) symmetry there are some subtleties at this stage. We describe how to handle such operators in Ref.~\cite{Hansen:2012tf}.} The three equations (\ref{eq:threeeqA}) may therefore be recast in the form
\begin{equation}
\vert (\mathrm{known\ number}) v_1 + (\mathrm{known\ number}) v_2 \vert = (\mathrm{known\ number}) \,.
\end{equation}
Any two equations constrain the values of the real parameters up to sign ambiguities which are lifted by the third equation.

\section{Conclusion}

We have presented the generalization of the L\"uscher quantization condition to an arbitrary number of strongly coupled, two-scalar-particle channels in a moving frame. We have also sketched the generalization of the Lellouch-L\"uscher formula, which gives decay amplitudes into these channels. Work is underway to further generalize this to final states with higher particle numbers.

\bibliographystyle{apsrev4-1}
\bibliography{ref}

\begin{thebibliography}{18}%
\makeatletter
\providecommand \@ifxundefined [1]{%
 \@ifx{#1\undefined}
}%
\providecommand \@ifnum [1]{%
 \ifnum #1\expandafter \@firstoftwo
 \else \expandafter \@secondoftwo
 \fi
}%
\providecommand \@ifx [1]{%
 \ifx #1\expandafter \@firstoftwo
 \else \expandafter \@secondoftwo
 \fi
}%
\providecommand \natexlab [1]{#1}%
\providecommand \enquote  [1]{``#1''}%
\providecommand \bibnamefont  [1]{#1}%
\providecommand \bibfnamefont [1]{#1}%
\providecommand \citenamefont [1]{#1}%
\providecommand \href@noop [0]{\@secondoftwo}%
\providecommand \href [0]{\begingroup \@sanitize@url \@href}%
\providecommand \@href[1]{\@@startlink{#1}\@@href}%
\providecommand \@@href[1]{\endgroup#1\@@endlink}%
\providecommand \@sanitize@url [0]{\catcode `\\12\catcode `\$12\catcode
  `\&12\catcode `\#12\catcode `\^12\catcode `\_12\catcode `\%12\relax}%
\providecommand \@@startlink[1]{}%
\providecommand \@@endlink[0]{}%
\providecommand \url  [0]{\begingroup\@sanitize@url \@url }%
\providecommand \@url [1]{\endgroup\@href {#1}{\urlprefix }}%
\providecommand \urlprefix  [0]{URL }%
\providecommand \Eprint [0]{\href }%
\providecommand \doibase [0]{http://dx.doi.org/}%
\providecommand \selectlanguage [0]{\@gobble}%
\providecommand \bibinfo  [0]{\@secondoftwo}%
\providecommand \bibfield  [0]{\@secondoftwo}%
\providecommand \translation [1]{[#1]}%
\providecommand \BibitemOpen [0]{}%
\providecommand \bibitemStop [0]{}%
\providecommand \bibitemNoStop [0]{.\EOS\space}%
\providecommand \EOS [0]{\spacefactor3000\relax}%
\providecommand \BibitemShut  [1]{\csname bibitem#1\endcsname}%
\let\auto@bib@innerbib\@empty
\bibitem [{\citenamefont {Luescher}(1986{\natexlab{a}})}]{Luescher:1986n1}%
  \BibitemOpen
  \bibfield  {author} {\bibinfo {author} {\bibfnamefont {M.}~\bibnamefont
  {L\"uscher}},\ }\href@noop {} {\bibfield  {journal} {\bibinfo  {journal}
  {Commun. Math. Phys.}\ }\textbf {\bibinfo {volume} {104}},\ \bibinfo {pages}
  {177} (\bibinfo {year} {1986}{\natexlab{a}})}\BibitemShut {NoStop}%
\bibitem [{\citenamefont {Luescher}(1986{\natexlab{b}})}]{Luescher:1986n2}%
  \BibitemOpen
  \bibfield  {author} {\bibinfo {author} {\bibfnamefont {M.}~\bibnamefont
  {L\"uscher}},\ }\href@noop {} {\bibfield  {journal} {\bibinfo  {journal}
  {Commun. Math. Phys.}\ }\textbf {\bibinfo {volume} {105}},\ \bibinfo {pages}
  {153} (\bibinfo {year} {1986}{\natexlab{b}})}\BibitemShut {NoStop}%
\bibitem [{\citenamefont {Luescher}(1991{\natexlab{a}})}]{Luescher:1991n1}%
  \BibitemOpen
  \bibfield  {author} {\bibinfo {author} {\bibfnamefont {M.}~\bibnamefont
  {L\"uscher}},\ }\href@noop {} {\bibfield  {journal} {\bibinfo  {journal}
  {Nucl. Phys.}\ }\textbf {\bibinfo {volume} {B354}},\ \bibinfo {pages} {531}
  (\bibinfo {year} {1991}{\natexlab{a}})}\BibitemShut {NoStop}%
\bibitem [{\citenamefont {Luescher}(1991{\natexlab{b}})}]{Luescher:1991n2}%
  \BibitemOpen
  \bibfield  {author} {\bibinfo {author} {\bibfnamefont {M.}~\bibnamefont
  {L\"uscher}},\ }\href@noop {} {\bibfield  {journal} {\bibinfo  {journal}
  {Nucl. Phys.}\ }\textbf {\bibinfo {volume} {B364}},\ \bibinfo {pages} {237}
  (\bibinfo {year} {1991}{\natexlab{b}})}\BibitemShut {NoStop}%
\bibitem [{\citenamefont {Lellouch}\ and\ \citenamefont
  {Luscher}(2001)}]{Lellouch:2000}%
  \BibitemOpen
  \bibfield  {author} {\bibinfo {author} {\bibfnamefont {L.}~\bibnamefont
  {Lellouch}}\ and\ \bibinfo {author} {\bibfnamefont {M.}~\bibnamefont
  {L\"uscher}},\ }\href@noop {} {\bibfield  {journal} {\bibinfo  {journal}
  {Commun.Math.Phys.}\ }\textbf {\bibinfo {volume} {219}},\ \bibinfo {pages}
  {31} (\bibinfo {year} {2001})}\BibitemShut {NoStop}%
\bibitem [{\citenamefont {Blum}\ \emph {et~al.}(2011)\citenamefont {Blum},
  \citenamefont {Boyle}, \citenamefont {Christ}, \citenamefont {Garron},
  \citenamefont {Goode} \emph {et~al.}}]{Blum:2011}%
  \BibitemOpen
  \bibfield  {author} {\bibinfo {author} {\bibfnamefont {T.}~\bibnamefont
  {Blum}},  \emph {et~al.},\ }\href {\doibase
  10.1103/PhysRevD.84.114503} {\bibfield  {journal} {\bibinfo  {journal}
  {Phys.Rev.}\ }\textbf {\bibinfo {volume} {D84}},\ \bibinfo {pages} {114503}
  (\bibinfo {year} {2011})},\ \Eprint {http://arxiv.org/abs/1106.2714}
  {arXiv:1106.2714 [hep-lat]} \BibitemShut {NoStop}%
\bibitem [{\citenamefont {Blum}\ \emph {et~al.}(2012)\citenamefont {Blum},
  \citenamefont {Boyle}, \citenamefont {Christ}, \citenamefont {Garron},
  \citenamefont {Goode} \emph {et~al.}}]{Blum:2011ng}%
  \BibitemOpen
  \bibfield  {author} {\bibinfo {author} {\bibfnamefont {T.}~\bibnamefont
  {Blum}},  \emph {et~al.},\ }\href {\doibase
  10.1103/PhysRevLett.108.141601} {\bibfield  {journal} {\bibinfo  {journal}
  {Phys.Rev.Lett.}\ }\textbf {\bibinfo {volume} {108}},\ \bibinfo {pages}
  {141601} (\bibinfo {year} {2012})},\ \Eprint {http://arxiv.org/abs/1111.1699}
  {arXiv:1111.1699 [hep-lat]} \BibitemShut {NoStop}%
\bibitem [{\citenamefont {Aaij}\ \emph {et~al.}(2012)\citenamefont {Aaij} \emph
  {et~al.}}]{LHCb:2011}%
  \BibitemOpen
  \bibfield  {author} {\bibinfo {author} {\bibfnamefont {R.}~\bibnamefont
  {Aaij}} \emph {et~al.} (\bibinfo {collaboration} {LHCb}),\
  }\href {\doibase 10.1103/PhysRevLett.108.111602} {\bibfield  {journal}
  {\bibinfo  {journal} {Phys. Rev. Lett.}\ }\textbf {\bibinfo {volume} {108}},\
  \bibinfo {pages} {111602} (\bibinfo {year} {2012})},\ \Eprint
  {http://arxiv.org/abs/1112.0938} {arXiv:1112.0938 [hep-ex]} \BibitemShut
  {NoStop}%
\bibitem [{\citenamefont {Aaltonen}\ \emph {et~al.}(2012)\citenamefont
  {Aaltonen} \emph {et~al.}}]{Collaboration:2012qw}%
  \BibitemOpen
  \bibfield  {author} {\bibinfo {author} {\bibfnamefont {T.}~\bibnamefont
  {Aaltonen}} \emph {et~al.} (\bibinfo {collaboration} {CDF}),\
  }\href {\doibase 10.1103/PhysRevLett.109.111801} {\bibfield  {journal}
  {\bibinfo  {journal} {Phys.Rev.Lett.}\ }\textbf {\bibinfo {volume} {109}},\
  \bibinfo {pages} {111801} (\bibinfo {year} {2012})},\ \Eprint
  {http://arxiv.org/abs/1207.2158} {arXiv:1207.2158 [hep-ex]} \BibitemShut
  {NoStop}%
\bibitem [{\citenamefont {Hansen}\ and\ \citenamefont
  {Sharpe}(2012)}]{Hansen:2012tf}%
  \BibitemOpen
  \bibfield  {author} {\bibinfo {author} {\bibfnamefont {M.~T.}\ \bibnamefont
  {Hansen}}\ and\ \bibinfo {author} {\bibfnamefont {S.~R.}\ \bibnamefont
  {Sharpe}},\ }\href {\doibase 10.1103/PhysRevD.86.016007} {\bibfield
  {journal} {\bibinfo  {journal} {Phys.Rev.}\ }\textbf {\bibinfo {volume}
  {D86}},\ \bibinfo {pages} {016007} (\bibinfo {year} {2012})},\ \Eprint
  {http://arxiv.org/abs/1204.0826} {arXiv:1204.0826 [hep-lat]} \BibitemShut
  {NoStop}%
\bibitem [{\citenamefont {Liu}\ \emph {et~al.}(2006)\citenamefont {Liu},
  \citenamefont {Feng},\ and\ \citenamefont {He}}]{Liu:2005kr}%
  \BibitemOpen
  \bibfield  {author} {\bibinfo {author} {\bibfnamefont {C.}~\bibnamefont
  {Liu}}, \bibinfo {author} {\bibfnamefont {X.}~\bibnamefont {Feng}}, \ and\
  \bibinfo {author} {\bibfnamefont {S.}~\bibnamefont {He}},\ }\href {\doibase
  10.1142/S0217751X06032150} {\bibfield  {journal} {\bibinfo  {journal}
  {Int.J.Mod.Phys.}\ }\textbf {\bibinfo {volume} {A21}},\ \bibinfo {pages}
  {847} (\bibinfo {year} {2006})},\ \Eprint
  {http://arxiv.org/abs/hep-lat/0508022} {arXiv:hep-lat/0508022 [hep-lat]}
  \BibitemShut {NoStop}%
\bibitem [{\citenamefont {Lage}\ \emph {et~al.}(2009)\citenamefont {Lage},
  \citenamefont {Meissner},\ and\ \citenamefont {Rusetsky}}]{Lage:2009}%
  \BibitemOpen
  \bibfield  {author} {\bibinfo {author} {\bibfnamefont {M.}~\bibnamefont
  {Lage}}, \bibinfo {author} {\bibfnamefont {U.-G.}\ \bibnamefont {Meissner}},
  \ and\ \bibinfo {author} {\bibfnamefont {A.}~\bibnamefont {Rusetsky}},\
  }\href {\doibase 10.1016/j.physletb.2009.10.055} {\bibfield  {journal}
  {\bibinfo  {journal} {Phys.Lett.}\ }\textbf {\bibinfo {volume} {B681}},\
  \bibinfo {pages} {439} (\bibinfo {year} {2009})},\ \Eprint
  {http://arxiv.org/abs/0905.0069} {arXiv:0905.0069 [hep-lat]} \BibitemShut
  {NoStop}%
\bibitem [{\citenamefont {Bernard}\ \emph {et~al.}(2011)\citenamefont
  {Bernard}, \citenamefont {Lage}, \citenamefont {Meissner},\ and\
  \citenamefont {Rusetsky}}]{Bernard:2010}%
  \BibitemOpen
  \bibfield  {author} {\bibinfo {author} {\bibfnamefont {V.}~\bibnamefont
  {Bernard}}, \bibinfo {author} {\bibfnamefont {M.}~\bibnamefont {Lage}},
  \bibinfo {author} {\bibfnamefont {U.-G.}\ \bibnamefont {Meissner}}, \ and\
  \bibinfo {author} {\bibfnamefont {A.}~\bibnamefont {Rusetsky}},\ }\href
  {\doibase 10.1007/JHEP01(2011)019} {\bibfield  {journal} {\bibinfo  {journal}
  {JHEP}\ }\textbf {\bibinfo {volume} {1101}},\ \bibinfo {pages} {019}
  (\bibinfo {year} {2011})},\ \Eprint {http://arxiv.org/abs/1010.6018}
  {arXiv:1010.6018 [hep-lat]} \BibitemShut {NoStop}%
\bibitem [{\citenamefont {Doring}\ \emph {et~al.}(2011)\citenamefont {Doring},
  \citenamefont {Meissner}, \citenamefont {Oset},\ and\ \citenamefont
  {Rusetsky}}]{Doering:2011}%
  \BibitemOpen
  \bibfield  {author} {\bibinfo {author} {\bibfnamefont {M.}~\bibnamefont
  {D\"oring}}, \bibinfo {author} {\bibfnamefont {U.-G.}\ \bibnamefont
  {Meissner}}, \bibinfo {author} {\bibfnamefont {E.}~\bibnamefont {Oset}}, \
  and\ \bibinfo {author} {\bibfnamefont {A.}~\bibnamefont {Rusetsky}},\ }\href
  {\doibase 10.1140/epja/i2011-11139-7} {\bibfield  {journal} {\bibinfo
  {journal} {Eur.Phys.J.}\ }\textbf {\bibinfo {volume} {A47}},\ \bibinfo
  {pages} {139} (\bibinfo {year} {2011})},\ \Eprint
  {http://arxiv.org/abs/1107.3988} {arXiv:1107.3988 [hep-lat]} \BibitemShut
  {NoStop}%
\bibitem [{\citenamefont {Aoki}\ \emph {et~al.}(2011)\citenamefont {Aoki} \emph
  {et~al.}}]{Aoki:2011}%
  \BibitemOpen
  \bibfield  {author} {\bibinfo {author} {\bibfnamefont {S.}~\bibnamefont
  {Aoki}} \emph {et~al.} (\bibinfo {collaboration} {HAL QCD}),\
  }\href@noop {} {\bibfield  {journal} {\bibinfo  {journal} {Proc.Japan Acad.}\
  }\textbf {\bibinfo {volume} {B87}},\ \bibinfo {pages} {509} (\bibinfo {year}
  {2011})},\ \Eprint {http://arxiv.org/abs/1106.2281} {arXiv:1106.2281
  [hep-lat]} \BibitemShut {NoStop}%
\bibitem [{\citenamefont {Kim}\ \emph {et~al.}(2005)\citenamefont {Kim},
  \citenamefont {Sachrajda},\ and\ \citenamefont {Sharpe}}]{Kim:2005}%
  \BibitemOpen
  \bibfield  {author} {\bibinfo {author} {\bibfnamefont {C.}~\bibnamefont
  {Kim}}, \bibinfo {author} {\bibfnamefont {C.}~\bibnamefont {Sachrajda}}, \
  and\ \bibinfo {author} {\bibfnamefont {S.~R.}\ \bibnamefont {Sharpe}},\
  }\href {\doibase 10.1016/j.nuclphysb.2005.08.029} {\bibfield  {journal}
  {\bibinfo  {journal} {Nucl.Phys.}\ }\textbf {\bibinfo {volume} {B727}},\
  \bibinfo {pages} {218} (\bibinfo {year} {2005})},\ \Eprint
  {http://arxiv.org/abs/hep-lat/0507006} {arXiv:hep-lat/0507006 [hep-lat]}
  \BibitemShut {NoStop}%
\bibitem [{\citenamefont {Davoudi}\ and\ \citenamefont
  {Savage}(2011)}]{Davoudi:2011md}%
  \BibitemOpen
  \bibfield  {author} {\bibinfo {author} {\bibfnamefont {Z.}~\bibnamefont
  {Davoudi}}\ and\ \bibinfo {author} {\bibfnamefont {M.~J.}\ \bibnamefont
  {Savage}},\ }\href {\doibase 10.1103/PhysRevD.84.114502} {\bibfield
  {journal} {\bibinfo  {journal} {Phys.Rev.}\ }\textbf {\bibinfo {volume}
  {D84}},\ \bibinfo {pages} {114502} (\bibinfo {year} {2011})},\ \Eprint
  {http://arxiv.org/abs/1108.5371} {arXiv:1108.5371 [hep-lat]} \BibitemShut
  {NoStop}%
\bibitem [{\citenamefont {Briceno}\ and\ \citenamefont
  {Davoudi}(2012)}]{Briceno:2012yi}%
  \BibitemOpen
  \bibfield  {author} {\bibinfo {author} {\bibfnamefont {R.~A.}\ \bibnamefont
  {Briceno}}\ and\ \bibinfo {author} {\bibfnamefont {Z.}~\bibnamefont
  {Davoudi}},\ }\href@noop {} {\  (\bibinfo {year} {2012})},\ \Eprint
  {http://arxiv.org/abs/1204.1110} {arXiv:1204.1110 [hep-lat]} \BibitemShut
  {NoStop}%
\end{thebibliography}%

\end{document}